\documentclass[useAMS,usenatbib]{mnras}

\usepackage{newtxtext,newtxmath}

\usepackage[T1]{fontenc}
\usepackage{ae,aecompl}

\usepackage{graphicx}
\usepackage{longtable}
\usepackage{float}
\usepackage[caption=false]{subfig}
\usepackage{amsmath}	
\usepackage{amssymb}	
\usepackage{amsfonts}
\usepackage{bm}
\title[ Periodicity in FRB 121102]{Possible periodic activity in the repeating FRB 121102}

\author[K. M. Rajwade et al.]{
K. M. Rajwade,$^{1}$\thanks{kaustubh.rajwade@manchester.ac.uk}
M. B. Mickaliger$^{1}$,
B. W. Stappers$^{1}$,
V. Morello$^{1}$,
D. Agarwal$^{2,3}$,
\newauthor
C. G. Bassa$^{4}$,
R. P. Breton$^{1}$,
M. Caleb$^{1}$,
A. Karastergiou$^{5,6,7}$,
E. F. Keane$^{8,1}$
\newauthor
and D. R. Lorimer$^{2,3}$
\\
$^{1}$Jodrell Bank Centre for Astrophysics, University of Manchester, Oxford Road, Manchester M13 9PL, UK\\
$^{2}$Department of Physics and Astronomy, West Virginia University, Morgantown, WV 26506,USA\\
$^{3}$Center for Gravitational Waves and Cosmology, West Virginia University, Morgantown, WV 26506, USA \\
$^{4}$ ASTRON, the Netherlands Institute for Radio Astronomy, Oude Hoogeveensedijk 4, 7991 PD Dwingeloo, The Netherlands \\
$^{5}$ Astrophysics, Denys Wilkinson building, University of Oxford, Keble Road, Oxford OX1 3RH, UK \\
$^{6}$Department of Physics and Electronics, Rhodes University, PO Box 94, Grahamstown 6140, South Africa\\
$^{7}$Physics Department, University of the Western Cape, Cape Town 7535, South Africa\\
$^{8}$ SKA Organisation, Jodrell Bank, Macclesfield SK11 9FT, UK
}

\date{Accepted XXX. Received YYY; in original form ZZZ}

\pubyear{\today}

\begin{document}
\label{firstpage}
\pagerange{\pageref{firstpage}--\pageref{lastpage}}
\maketitle

\begin{abstract}
The discovery that at least some Fast Radio Bursts (FRBs) repeat has ruled out cataclysmic events as the progenitors of these particular bursts. FRB~121102 is the most well-studied repeating FRB but despite extensive monitoring of the source, no underlying pattern in the repetition has previously been identified. Here, we present the results from a radio monitoring campaign of FRB~121102 using the 76-m Lovell telescope. Using the pulses detected in the Lovell data along with pulses from the literature, we report a detection of periodic behaviour of the source over the span of five years of data. We predict that the source is currently `off' and that it should turn `on' for the approximate MJD range $59002-59089$ (2020-06-02 to 2020-08-28). This result, along with the recent detection of periodicity from another repeating FRB, highlights the need for long-term monitoring of repeating FRBs at a high cadence. Using simulations, we show that one needs at least 100 hours of telescope time to follow-up repeating FRBs at a cadence of 0.5--3 days to detect periodicities in the range of 10--150 days. If the period is real, it shows that repeating FRBs can have a large range in their activity periods that might be difficult to reconcile with neutron star precession models.


\end{abstract}




\section{Introduction}
Fast Radio Bursts (FRBs) are bright radio pulses that last for no more than a few milliseconds~\citep{Lorimer,Thornton}. While their nature is still a mystery, we know they are extragalactic on account of their anomalously high dispersion measures as well as the measured redshifts of the host galaxies of localized FRBs~\citep{SriHarsh,Ravi,bannister2019}. Although subject to large variance at lower redshifts~\citep{Masui}, the DM acts as a reasonable proxy for distance on cosmological scales~\citep{Keane2018}. In spite of detections only at radio wavelengths, the data not only contain information on the intergalactic medium but also about the progenitor and its local environment~\citep{Masui}. 

To date, more than one hundred FRBs have been published~\citep{frbcat}, yet only some of these have so far been observed to repeat \citep{nat_spitler, nat_Shannon, Andersen, Kumar} and there is no clear evidence favouring a specific progenitor model. The first repeater, FRB 121102, was discovered in 2014~\citep{Spitler2014} though its repeating nature was not revealed until 2016~\citep{Spitler2016}. This discovery was crucial as it implied that not all FRB progenitors were of cataclysmic origin. Since then, 19 more repeaters have been discovered~\citep{fonseca2020,CHIME2019c,Kumar}. While the new discoveries suggest the possibility of multiple populations of FRBs, a lack of urgent follow-up and monitoring of all known FRBs precludes a definitive conclusion. Of all the repeating sources, FRB~121102 has been studied extensively across a broad range of radio frequencies from 600 MHz~\citep{Josephy} to 8 GHz~\citep{Gajjar}. Though numerous pulses have been detected to date, no underlying pattern has been discovered so far. The shortest separations between two apparently distinctive pulses are 26~ms~\citep{Gourdji}, 34~ms~\citep{Hardy} and 37~ms~\citep{Scholz}. 
The recent discovery of periodic activity from FRB~180906.J0158+65~\citep{CHIME2020a} has rekindled interest in this question and leads us to wonder whether all repeating FRBs show this kind of behaviour. The 16.35-day periodicity in FRB~180906.J0158+65 has led to models being invoked such as; binary orbits to explain the observed periodic behaviour~\citep{Lyu2020,zhang2020} while some authors have proposed a precession of flaring, highly magnetized neutron stars~\citep{Yuri2020,Zanazzi2020}. If true, it will provide a vital clue into the origins of these mysterious bursts. In this paper, we present the results of a long-term monitoring campaign of FRB~121102 using the 76-m Lovell telescope (LT) located at the Jodrell Bank Observatory. The observing campaign is presented in \S~2. We then describe our search for periodic activity in \S~3. We discuss the results obtained in \S~4 before providing concluding remarks in \S~5.

\begin{figure}
	\includegraphics[width=\columnwidth]{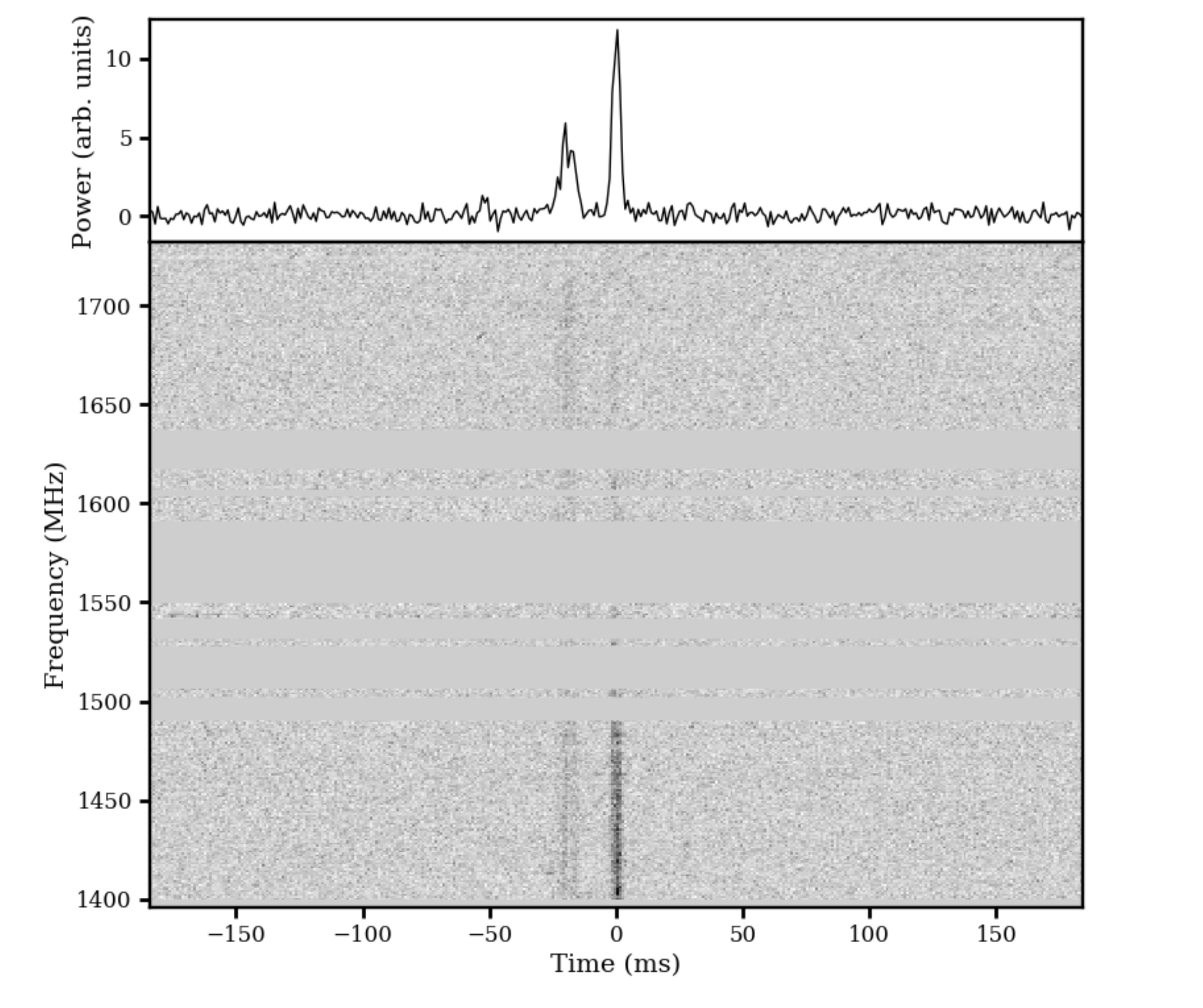}
    \caption{Dynamic spectrum of a pulse of FRB~121102 detected by the Lovell telescope. Some of the frequency channels have been flagged as they were corrupted by strong RFI. One can clearly see a pre-cursor to the main pulse that is separated by $\sim$17 milliseconds.}
    \label{fig:dyn_spec}
\end{figure}

\section{Observations \& data processing}
\begin{figure*}
    \centering
    \includegraphics[width=17cm]{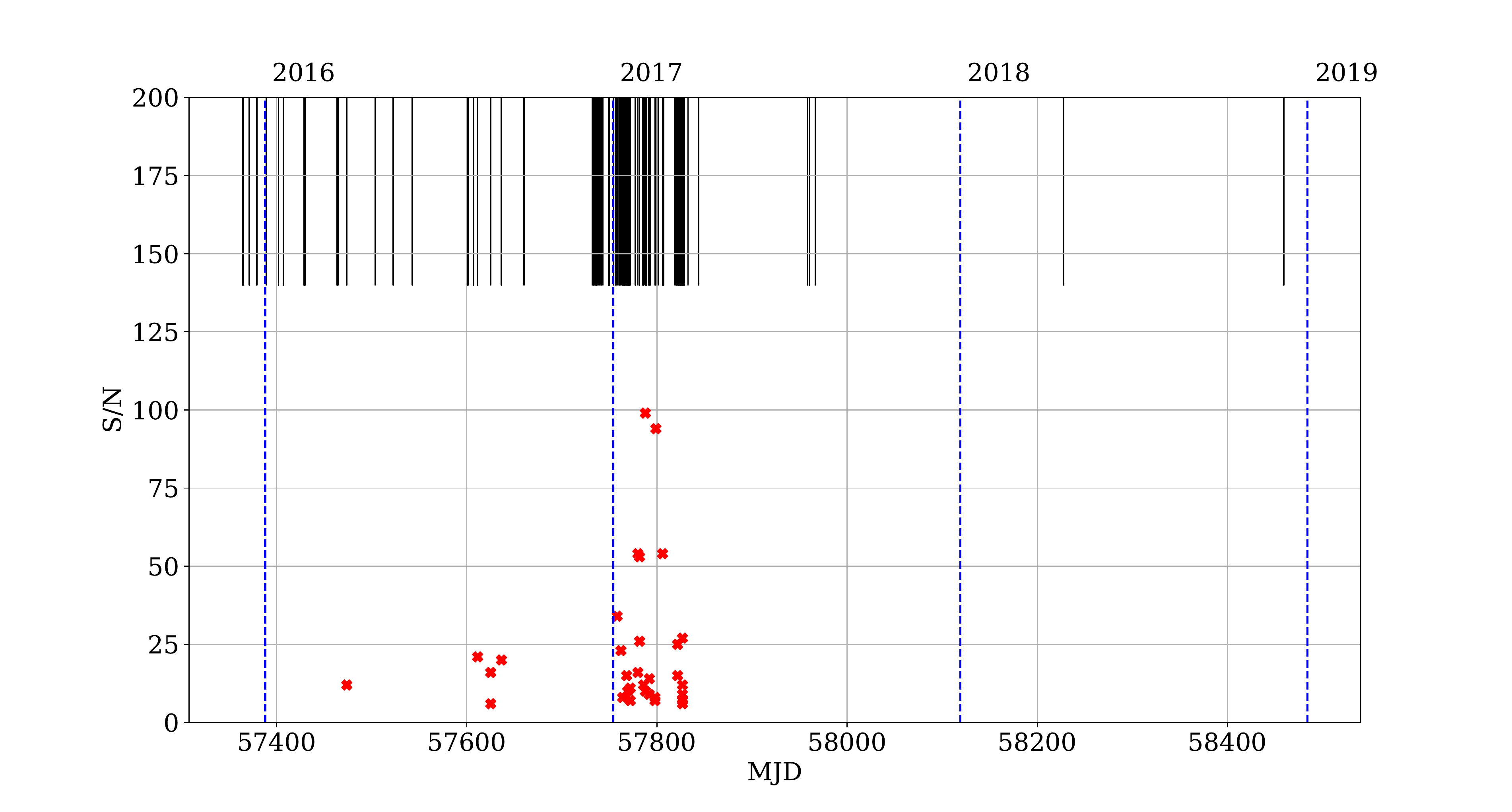}
    \includegraphics[width=17cm]{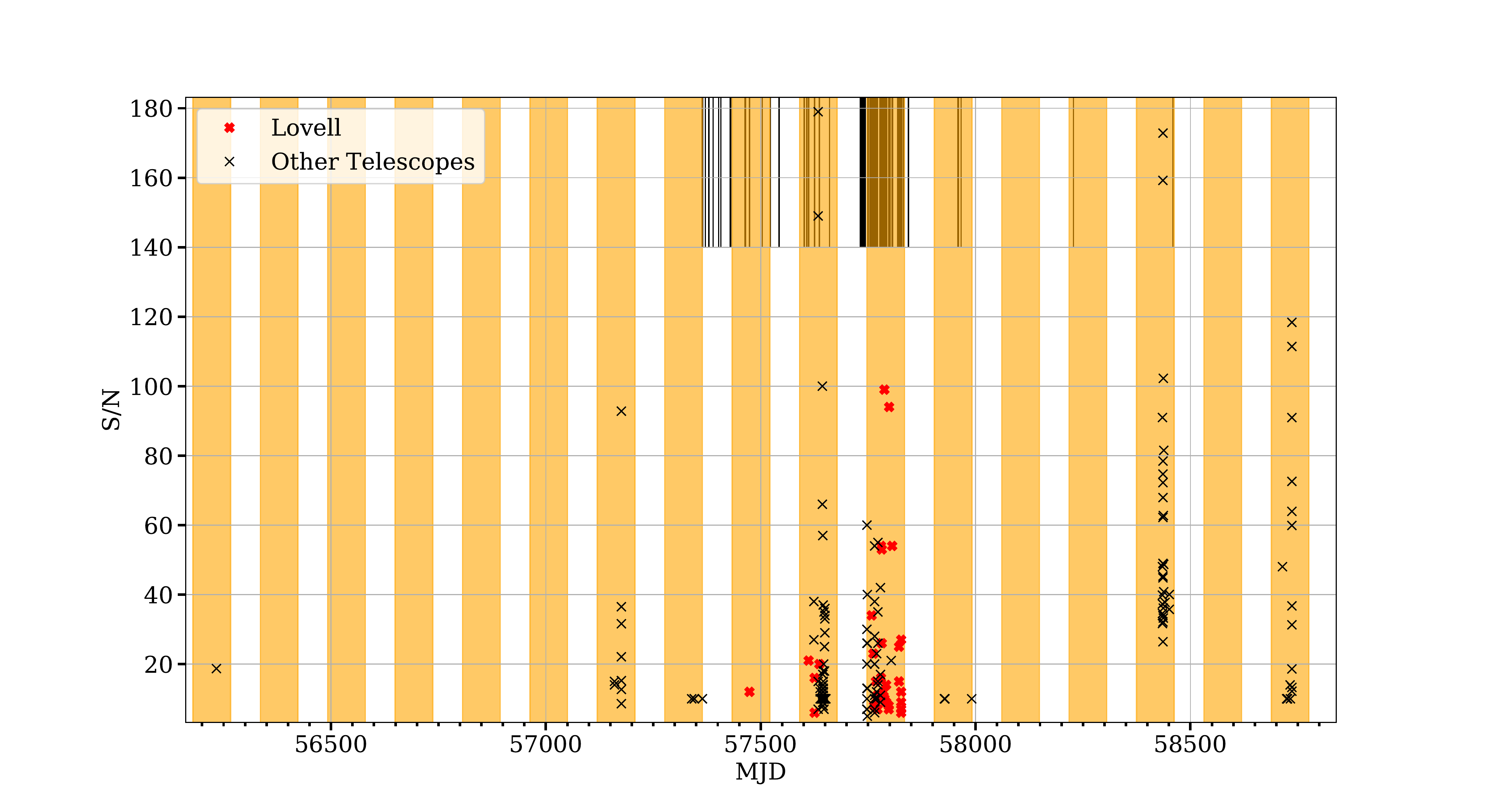}
    \caption{\textbf{Top panel}: S/N ratio versus MJD for all FRB~121102 detections in the LT monitoring campaign. Black vertical lines denote the LT observation dates over the span of the campaign. Blue dotted lines correspond to 1${\rm st}$ of January of the year shown on top of the panel. \textbf{Bottom panel}: detection MJDs as a function of S/N for pulses of FRB 121102 from the published literature and this campaign. The black vertical lines denote the observation during the monitoring campaign by the LT. The orange shaded region shows the best detected period of activity that is phased to the reference MJD of 58200 and then extrapolated over the entire span. For detections where the S/N was not reported in the literature, we have used a S/N of 10. The references for all pulses taken from the literature and used here are presented in Table~\ref{tab:alldets}.}
    \label{fig:R1_allobs}
\end{figure*}

Since the discovery of repeating pulses from FRB~121102, it was followed up on a pseudo-regular basis using the LT. Starting from MJD 57363, the source was followed up on a near-weekly cadence, with some daily observations, until MJD 57723. From that point, it was observed nearly every day through MJD 57843. After that, there were a few sparse observations until 58483. The cadence of the monitoring campaign was non-uniform, the observations are interspersed with large gaps due telescope maintenance. The top panel of Figure~\ref{fig:R1_allobs} shows the cadence of observations over the last 4 years. Details of all observations with LT are shown in Table~\ref{tab:LTobs}.

\begin{table}
\centering
\begin{tabular}{l|l|l|l}
 ID       &    MJD Start & MJD End & No. of detections \\                 
\hline\hline
0   & 57363.9001968 & 57363.9842014 & 0 \\
1   & 57365.0012963  & 57365.0853009 & 0 \\
2   & 57371.1676968  & 57371.2515278 & 0 \\
3   & 57379.2055208 & 57379.289537 & 0 \\
4   & 57389.0533796 & 57389.1374074 & 0 \\
5   & 57402.0379051 & 57402.1219213 & 0 \\
6   & 57407.1136111 & 57407.1976273 & 0 \\
7   & 57428.8283333 & 57428.9998611 & 0 \\
8   & 57429.0005556 & 57429.9993519 & 0 \\
9   & 57430.0000463 & 57430.079213 & 0 \\
10  & 57463.5381713 & 57463.6265741 & 0 \\
\hline
\end{tabular}
\caption{Start and end MJDs for all observations with the LT and the number of detections in each observing session. The full table can be found in the the online supplementary materials.}
\label{tab:LTobs}
\end{table}

For each observation, a polyphase filter coarsely channelized a 400~MHz band into 25 subbands of 16~MHz each using a ROACH-based backend~\citep{bassa+16}. Each 16~MHz subband was further channelized into $32 \times 0.5$~MHz channels using \texttt{digifil} from the \texttt{dspsr} software suite~\citep{2011VanStraten}, and downsampled to a sampling time of 256~$\upmu$s. The 800 total channels, spanning 400~MHz, were then combined  in frequency. After MJD 57729, all observations (75\% of data reported here) had a bandwidth of 336 MHz, to mitigate the effect of radio-frequency interference (RFI) on the data. We also masked frequency channels in the data containing narrow-band RFI.

No other RFI mitigation algorithm was used to massage the data. We searched these data using the single-pulse-search software package \textsc{heimdall}\footnote{\url{https://sourceforge.net/projects/heimdall-astro/}} that searches for single pulses over a timeseries generated for a range of trail DMs using a brute-force dedispersion algorithm. We used a DM range of 0 to 800~pc~cm$^{-3}$ and searched over widths ranging from 256~$\upmu$s up to 32~ms. Candidates from \textsc{heimdall} were classified with the \textsc{FETCH} machine-learning candidate classifier~\citep{agarwal2019}, and all candidates classified as astrophysical pulses, with a signal-to-noise ratio (S/N) greater than 8, were viewed by eye to verify they were real, astrophysical pulses from FRB~121102. From this analysis, we detected 25 pulses in the data. To look for fainter pulses, we visually inspected all candidates down to a S/N of 6; 7 more pulses were found. \textsc{FETCH} misclassified five of the seven low S/N pulses as the neural network is not trained on any pulses with S/N less than 8. The dynamic spectrum of one of the pulses detected with the Lovell Telescope is shown in Figure~\ref{fig:dyn_spec}. For each pulse, the cleaned data were dedispersed at the S/N optimized DM. We know that the true DM of this source is different owing to structure in the radio emission that varies over time and frequency~\citep{Hessels}. Since structure analysis is not the focus of this paper, we decided to dedisperse the pulses to maximise the S/N, which will mean that the difference in these DMs for each of the Lovell pulses mostly arises from the presence or absence of frequency drifting observed in the components of the pulses and not from changes in the total electron column density. The resulting timeseries were convolved with a series of Gaussian templates over a range of widths using a python based package \textsc{spyden}~\footnote{\url{https://bitbucket.org/vmorello/spyden/}} to obtain the best-fit S/N and width for each pulse. Then, we computed the fluence for each pulse using the radiometer equation~\citep{handbook2004}. For a given S/N and width, $W$, the fluence,
\begin{equation}
     \mathcal{F} = {\rm S/N}~\frac{G~T_{\rm sys}\sqrt{W}}{\sqrt{n_{p}~\Delta\nu}},
\end{equation}
where $G$ is the telescope gain ($G$ $\simeq$ 0.9) in units of K~Jy$^{-1}$, $T_{\rm sys}$ is the system temperature that is the summation of the receiver temperature and the sky temperature at the centre frequency of the receiver in Kelvin, $n_{p}=2$ is the number of polarizations to be summed and $\Delta \nu$ is the bandwidth in Hz. The calculated parameters for each pulse are presented in Table~\ref{tab:par}. 

\begin{table}
    \centering
	\begin{tabular}{l|l|l|l|l|l}
	    \hline
	     ID & Topocentric MJD & Fluence & Width & S/N & DM\\
	    \hline
	     &  & Jy~ms & ms &  \\
		\hline
        1 & 57473.846689 & 1.29(10)  & 3.5 & 12 & 559.5  \\
        2 & 57611.452953 & 1.50(7)  & 1.6 & 21 & 560.5\\
        3 & 57625.246712 & 0.40(6)  & 1.2 & 6 & 559.5\\
        4 & 57625.247667 & 1.63(10)  & 3.2 & 16 & 562.4 \\
        5 & 57636.489603 & 2.16(10)  & 3.6  & 20 & 571.2 \\
        6 & 57758.162612 & 5.23(15)  & 7.2  & 34 & 557.5 \\
        7 & 57762.155348 & 1.74(7)  & 1.7  & 23 & 559.8 \\
        8 & 57763.975657 & 0.57(7)  & 1.6  & 8  & 559.8\\
        9 & 57768.159477 & 1.28(8)  & 2.2 & 15 & 562.1\\
        10& 57769.143333 & 0.85(8)  & 2.2 & 10 & 558.6\\
        11& 57771.954804 &  0.88(8)  & 1.9  & 11 & 558.6 \\
        12& 57771.958773 & 0.59(8)  & 2.2 & 7 & 560.9 \\
        13& 57779.957530 & 5.84(10)  & 3.6 & 54 & 562.1 \\
        14& 57779.978393 & 1.21(7)  & 1.7 & 16 & 563.3 \\
        15& 57781.770722 & 4.80(9)  & 2.5 & 53 & 562.1 \\
        16& 57781.771322 & 3.35(12)  &5.0 & 26 & 569.1\\
        17& 57785.973376 & 1.15(9)  &2.8  & 12 & 560.9\\
        18& 57787.822048 & 0.85(8)  & 2.2 & 10 & 562.1\\
        19& 57787.844951 & 11.34(11) &3.9  & 99 & 563.3\\
        20& 57791.942210 & 0.72(8) & 1.9 & 9 & 562.1\\
        21& 57791.946845 & 1.12(9) & 1.9 & 14 & 569.1 \\
        22& 57797.926712 & 0.81(10) & 3.2 & 8 & 560.9\\
        23& 57797.930046  & 0.53(7) & 1.7 & 7 & 565.6\\
        24& 57798.872124 & 11.42(12) & 4.5  & 94 & 565.6 \\
        25& 57805.959486 & 4.89(9)  & 2.5 & 54 & 559.8\\
        26& 57821.785328 & 1.28(8) & 2.2 & 15 & 563.3\\
        27& 57821.789488 & 3.22(12)  & 5.0 & 25 & 571.5 \\
        28& 57826.841596 & 1.02(8) & 2.2 & 12 & 563.3\\
        29& 57826.845833 & 1.93(7) & 1.6 & 27 & 560.9\\
        30& 57826.851906 & 0.68(7) & 1.7 & 9 & 560.9\\
        31& 57826.862280 & 0.44(6) & 1.2 & 7 & 562.1\\
        32& 57826.865941 & 0.45(7) & 1.7 & 6 & 560.9\\
        \hline
	\end{tabular}
	\caption{Observed and derived parameters of the detected pulses from FRB 121102 during the LT monitoring campaign. The values in the parenthesis indicate the 1-$\sigma$ uncertainty on the least significant digit(s). The DMs correspond to the DM at which the S/N was maximum for each pulse. The topocentric MJDs correspond to the MJD of the burst at the highest frequency of the LT  (1712~MHz).}
	\label{tab:par}
\end{table}

\section{Periodicity}

Table~\ref{tab:par} shows the observed parameters of the detected FRBs in the monitoring campaign. The time-span of more than two years enabled us to study in detail the long term emission variability of FRB~121101. The top panel of Figure~\ref{fig:R1_allobs} shows the LT detections along with the observing dates over the entire campaign. Visually, we noticed a pattern in the detection of pulses from FRB~121102. To make sure that we are not biased by unevenly sampled observations, we ran a two sample Wald-Wolfovitz runs test~\citep{Alhakim2008} on the LT sample of pulses. This test evaluates whether a given sequence of binomial outcomes is likely to be drawn from a random distribution. Here, a run is defined as a sequence of same consecutive outcomes. Hence, for a given sequence of events with two outcomes, the test statistic,
\begin{equation}
    Z = \frac{R -\Bar{R}}{S_{R}},
\end{equation}
where $R$ is the observed number of runs, the expected number of runs,
\begin{equation}
    \bar{R} = \frac{2~n_{1}n_{2}}{n_{1} + n_{2}} + 1,
\end{equation}
and the standard deviation of runs,
\begin{equation}
    S_{R} = \frac{2 n_{1}n_{2} (2n_{1}n_{2} - n_{1} - n_{2})}{(n_{1} + n_{2})^{2} (n_{1} + n_{2} - 1)},
\end{equation}
in which $n_{1}$ and $n_{2}$ are the total number of runs for each outcome.
$Z$ can then be tested against the null hypothesis by comparing its value with the normal table for a given significance. In our case, we assigned observations where we had detections as ``P'' and non-detections as ``N'' that generated a sequence over the entire observing campaign. We found that $Z=-2.08$. This  rejects the null hypothesis at a 96\% significance level and shows that the detection sequence is an unlikely outcome from a purely random sequence. However, we note that this does not mean that there is an underlying periodicity in the activity of FRB~121102 as a two sample test only confirms whether there is a dependence between the two outcomes of the sequence.

\begin{table}
\centering
\begin{tabular}{l|l|l}
 ID       &    MJD & Reference  \\                 
\hline\hline
0   & 56233.282837007995 & Spitler et al. 2016 \\
1   & 57159.737600835    & Spitler et al. 2016 \\
2   & 57159.744223619    & Spitler et al. 2016 \\
3   & 57175.693143232005 & Spitler et al. 2016 \\
4   & 57175.699727825995 & Spitler et al. 2016 \\
5   & 57175.742576706    & Spitler et al. 2016 \\
6   & 57175.742839344006 & Spitler et al. 2016 \\
7   & 57175.743510388    & Spitler et al. 2016 \\
8   & 57175.745665832    & Spitler et al. 2016 \\
9   & 57175.747624851    & Spitler et al. 2016 \\
10  & 57175.748287265    & Spitler et al. 2016 \\
\hline
\end{tabular}
\caption{MJDs of the first 10 published pulses of FRB~121102 used in this paper. Full table can be found in the online supplementary materials}
\label{tab:alldets}
\end{table}

\begin{figure*}
    \centering
    \includegraphics[width=\textwidth]{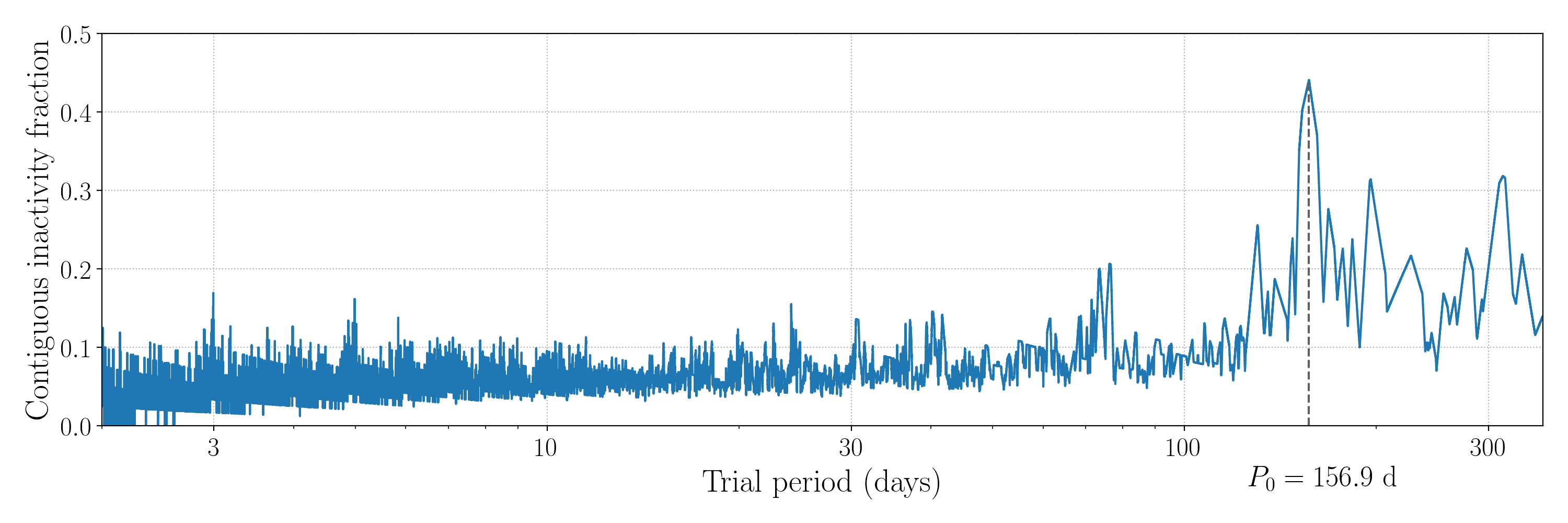}
    \caption{Periodogram obtained by running a Fast Folding Algorithm (FFA) on an evenly sampled, high time-resolution histogram of the detected pulse MJDs. The folded profiles produced by the FFA were evaluated by the length (relative to the trial period) of the longest contiguous phase region \textit{without} detectable activity. At the most significant trial period, $P_0 = 156.9$ days, the source is active only for a contiguous 56\% of a hypothetical cycle.}
    \label{fig:ffa_periodogram}
\end{figure*}

\begin{figure}
	\includegraphics[width=\columnwidth]{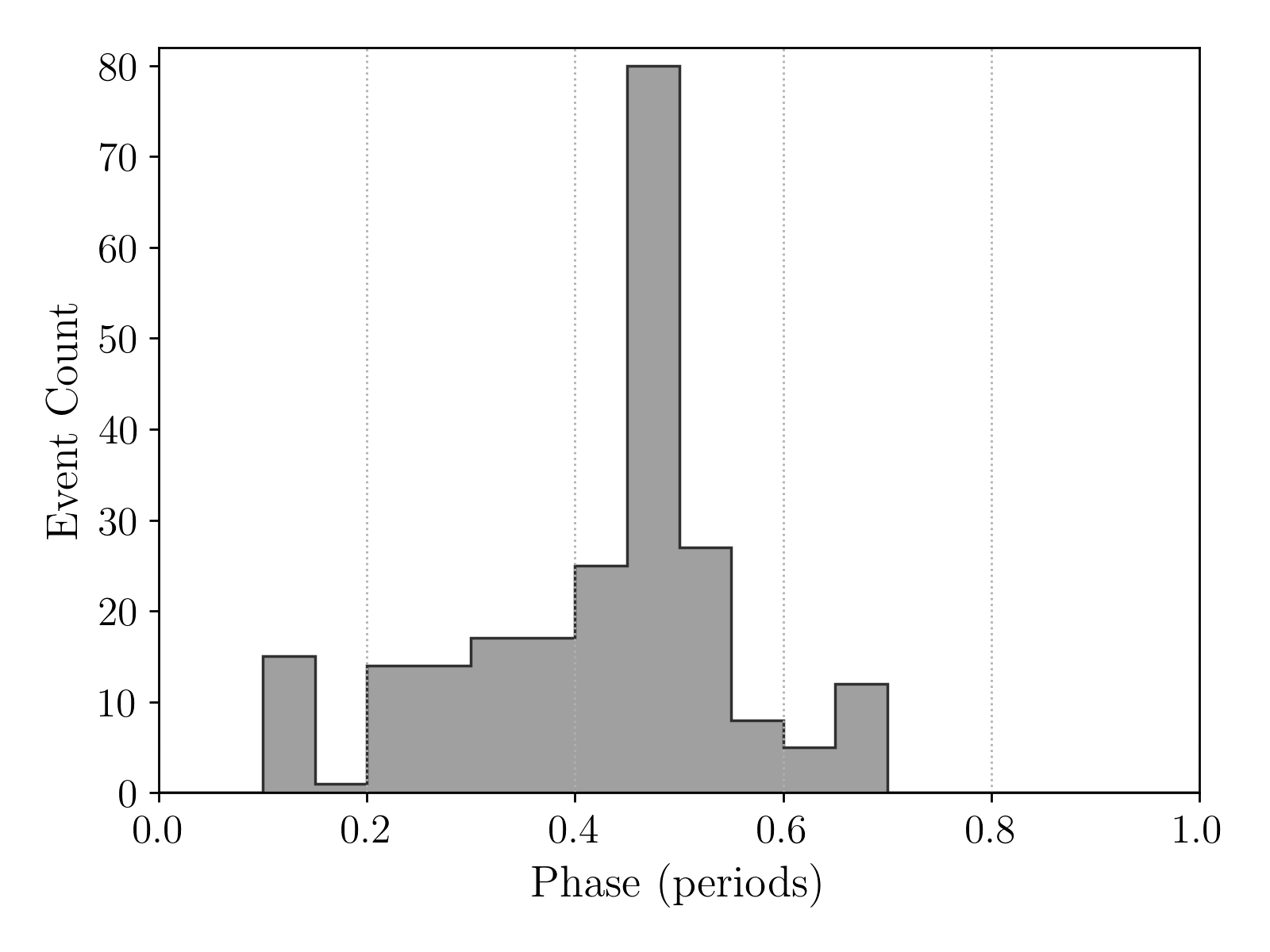}
    \caption{Detected pulse MJDs folded at the best-fit period of $P_0 = 157$ days (). A phase of zero corresponds to the reference MJD $t_{\mathrm{ref}} = 58200$. Note that the peak at the phase of 0.5 may not be real as not all phases of the activity period have been sampled uniformly.}
    \label{fig:folded_pulse_mjds}
\end{figure}

To confirm the periodic behaviour in the activity of FRB 121102, we first tried a Lomb--Scargle periodogram~\citep{Scargle1982}. Since the LT observations are spread along a long baseline and tend to be densely sampled closer to periods of activity, a periodogram of the resulting timeseries was biased. One needs to sample multiple active and inactive periods to get a correct period from the periodogram even though the sampling is non-uniform~\citep[see][for more details]{vander2018}. To overcome this, we used a Fast Folding Algorithm (FFA) to search for periodicity in the activity of the source. The FFA is designed to search for periodic pulsar signals in time series data, and provides the highest possible period resolution for that purpose \citep{Staelin1969}. To make the algorithm applicable to our data set, we first binned the list of detected pulse MJDs available in the literature, 215 MJDs in total from~\cite{Spitler2014,Scholz,scholz2017,Hardy, Gourdji,Spitler2016,Chatterjee,Marcote,law2017,spitler2018,gajjar2018,Hessels,oostrum2019} (see Table~\ref{tab:alldets} for more details) and this paper, into a histogram with a time resolution of 0.05 days. We don't use the most recent active phase that was reported by multiple telescopes (MJD $>$ 58500)~\citep{li2019, Pearlman2019, CalebATel}. Using an FFA implementation\footnote{\url{https://github.com/v-morello/riptide}}~\citep{Morello2020}, we then phase-coherently folded these data at all distinguishable trial periods between 2 and 365 days, which generated sets of profiles representing source activity as a function of phase for all trial periods. When the FFA is used for pulsar searching, the folded profiles it produces are usually tested for significance with sets of matched filters reproducing an expected pulse shape, or a $\chi^2$ test. Here, however, such methods would be ineffective as most of the detections are concentrated within short time spans and therefore tend to be folded in only a few distinct phase bins regardless of trial period. We therefore used a modified metric: in each fold trial, we measured the length of the longest contiguous phase segment (in units of period) without any source activity. Higher values denote that the activity of the source is concentrated within a smaller phase window, which indicates a periodic activity pattern. The fraction of source inactivity as a function of period is plotted in Figure ~\ref{fig:ffa_periodogram}. We find that for a trial period $P_0 = 157 \pm 7$ days, the source remains inactive for a contiguous 44\% of the time within each putative cycle. The behaviour of the inactivity metric as a function of trial period cannot be modeled analytically which precludes deriving a mathematically rigorous uncertainty on $P_0$, and thus the error bars provided correspond to the full width at half-maximum of the periodogram peak. We produced an activity profile of the source by folding the MJDs of the detected pulses at the best-fit period $P_0 = 157$~days, which is displayed in Figure~\ref{fig:folded_pulse_mjds}. Using the detected period and a duty cycle of 56$\%$ we extrapolated the activity period over the span of four years of observations including all published detections of FRB~121102 to date (see Table~\ref{tab:alldets} for details of all detections) and the results are presented in Figure~\ref{fig:R1_allobs}. One can see that the activity period aligns very nicely with the until now excluded detections by the MeerKAT telescope (Caleb et al., in prep.), the FAST telescope~\citep{li2019} and the Deep Space Network~\citep{Pearlman2019}. Hence, all the evidence presented here suggests that this is the most likely activity period of FRB~121102.

\section{Discussion}
\subsection{Periodic Activity?}

Here, we discuss the significance of the detected periodicity in FRB~121102.
Due to the sparse and uneven observing coverage of the whole time span considered (Fig.~\ref{fig:R1_allobs}), we cannot reasonably assume that the pulse detection dates are uniformly distributed in phase under the null hypothesis (i.e., the source exhibits no periodic activity pattern) for a period $P_0 = 157$ days. To estimate that distribution, an exhaustive list of the start and end dates of all attempted observations would be required, but is not available since typically, only detected pulse MJDs are published in the literature. The statistical significance of our detected periodicity thus cannot be rigorously estimated with the data currently available, and should be treated circumspectly as it may result from a chance alignment between the time ranges where no observations have been made. We acknowledge that bootstrapping the available detections is a possible option~\citep{EfroTibs93}. In this method, one can re-sample the arrival times of the bursts from a uniform distribution and run a periodicity search on the obtained arrival times. By running multiple realizations of the same, one can obtain a probability of detecting the period by chance. Though bootstrapping can give some sort of a significance for the detected peak, the main caveat of this method is the assumption that all the observations conducted in a given time period are randomly distributed over the entire time period. This is not true with follow-ups of repeating sources as telescopes tend to observe these sources with denser cadence when there is a previously known detection. We also note that if the periodic activity in FRB~121102 is in any way similar to FRB180916.J0158+65, one would expect the source to not emit in every single active phase. This can also result in reduction in the significance of detection of periodicity.


Our best-fit parameters suggest that the next two activity periods should occur in the MJD ranges 59002$-$59089 (2020-06-02 to 2020-08-28) and 59158$-$59246 (2020-11-05 to 2021-02-01). We particularly encourage further observations during the predicted quiescence period in-between, as they could falsify our periodicity claim. A confirmation will require extending the baseline of observations, preferably with a regular cadence. How to optimally space observations to search for, or confirm periodicity of a repeating source is a question that deserves further examination. In essence, a large number of cycles need to be sampled before any proper statistical analysis on the significance of detection can be performed. 

If the detected period is astrophysical in origin, it has implications on the possible progenitors of repeating FRBs.~\cite{CHIME2020a} have invoked orbital motion to cause such periodicities. If we consider now also orbital motion to be the cause of the observed periodicity in FRB~121102, the large range in the observed periods (16--160 days) can constrain the possible binary systems. High-mass X-ray binaries are systems with a neutron star in an orbit with a massive O/B star. HMXBs in our Galaxy and the Small Magellanic Cloud have a large range of orbital periods, ranging from few tens to hundreds of days~\citep[see][for more details]{liu2006}.~\cite{zhang2020} propose a model where the magnetized neutron star is combed by the highly energetic wind of the secondary star. Massive stars in HMXB systems tend to possess energetic winds for this scenario to be feasible. On the other hand, binaries where the donor star fills the Roche lobe of the system have much shorter periods ($<10$~days) and are unlikely to be possible progenitors.
Other progenitor models invoke precessing  neutron stars or young flaring magnetars~\citep{Yuri2020,Zanazzi2020}. The authors of these studies expect the timescale of precession to be of the order of weeks though larger precession periods (a few months) would be harder to explain as the internal magnetic field would have to be lower by at least a factor of 3 compared to the expected internal fields in young magnetars and will have implications on the observed burst energies from these sources~\citep{Yuri2020}. To draw any inferences about the origin of this repeating class of FRBs, regular monitoring of such sources is imperative along with more discoveries of periodic FRBs and a systematic approach to following up known repeaters with existing instruments can achieve this goal.

\subsection{Follow-up Strategies}
\begin{figure*}
\subfloat{\includegraphics[scale=0.33]{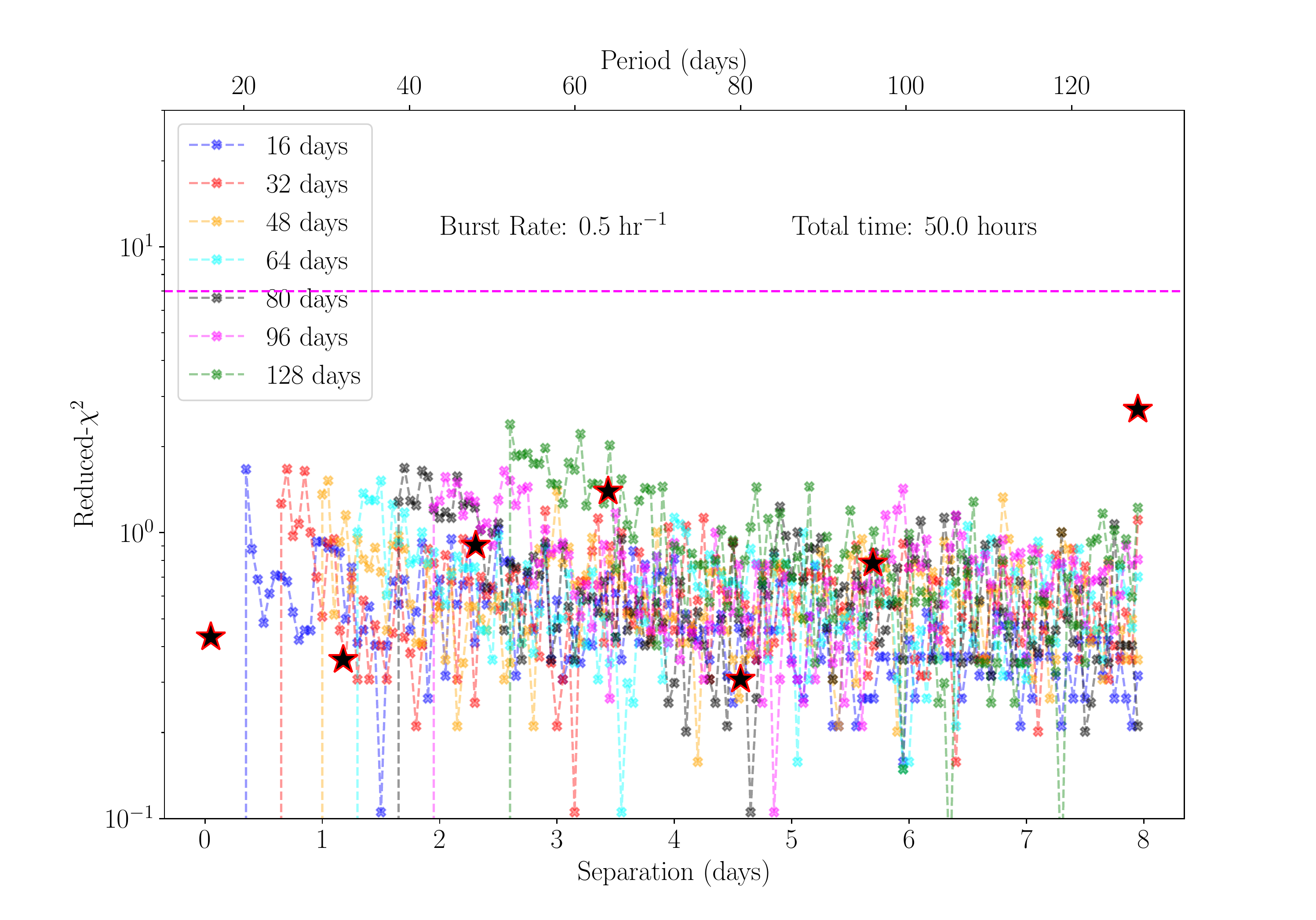}} 
\subfloat{\includegraphics[scale=0.33]{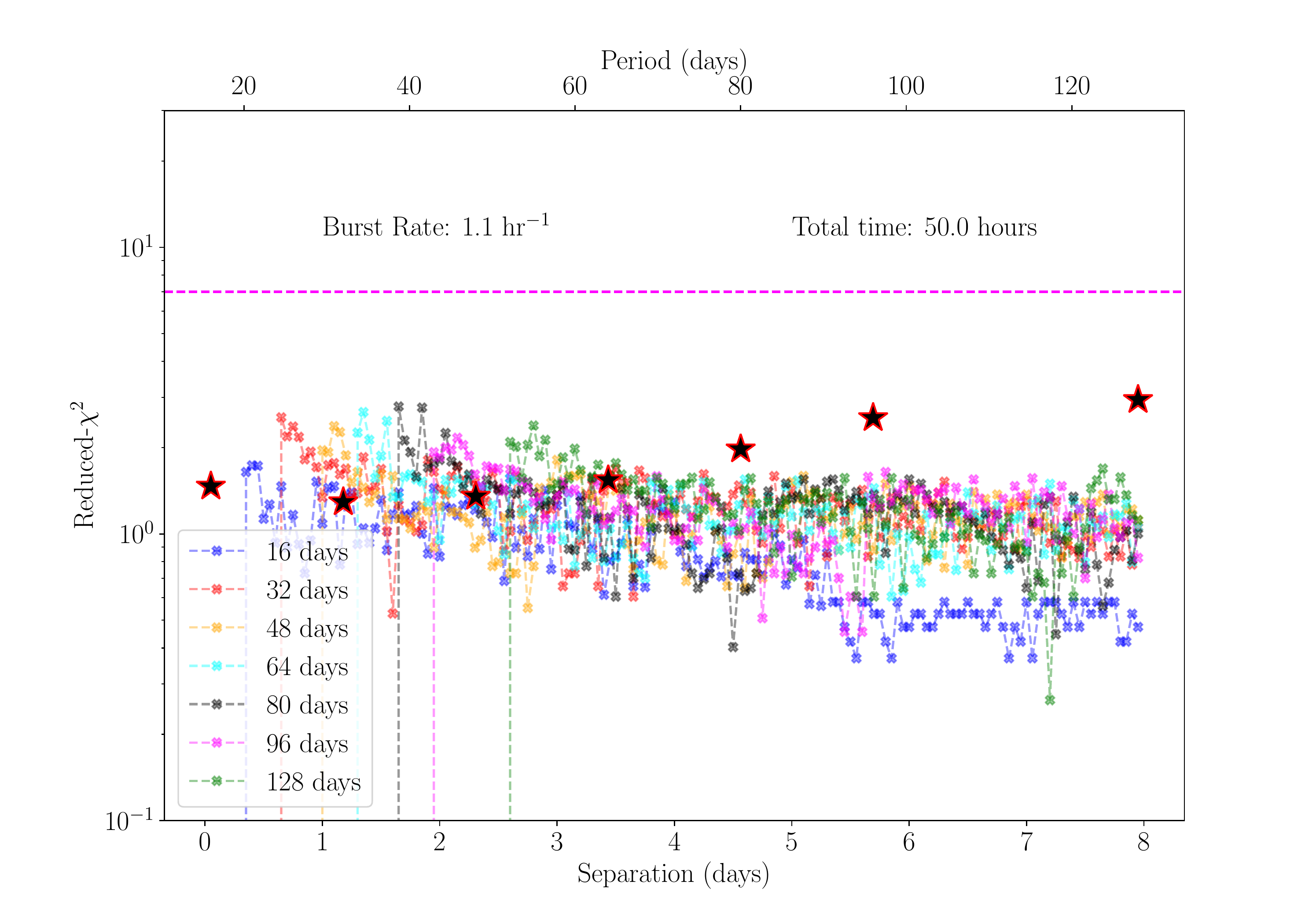}}\\
\subfloat{\includegraphics[scale=0.33]{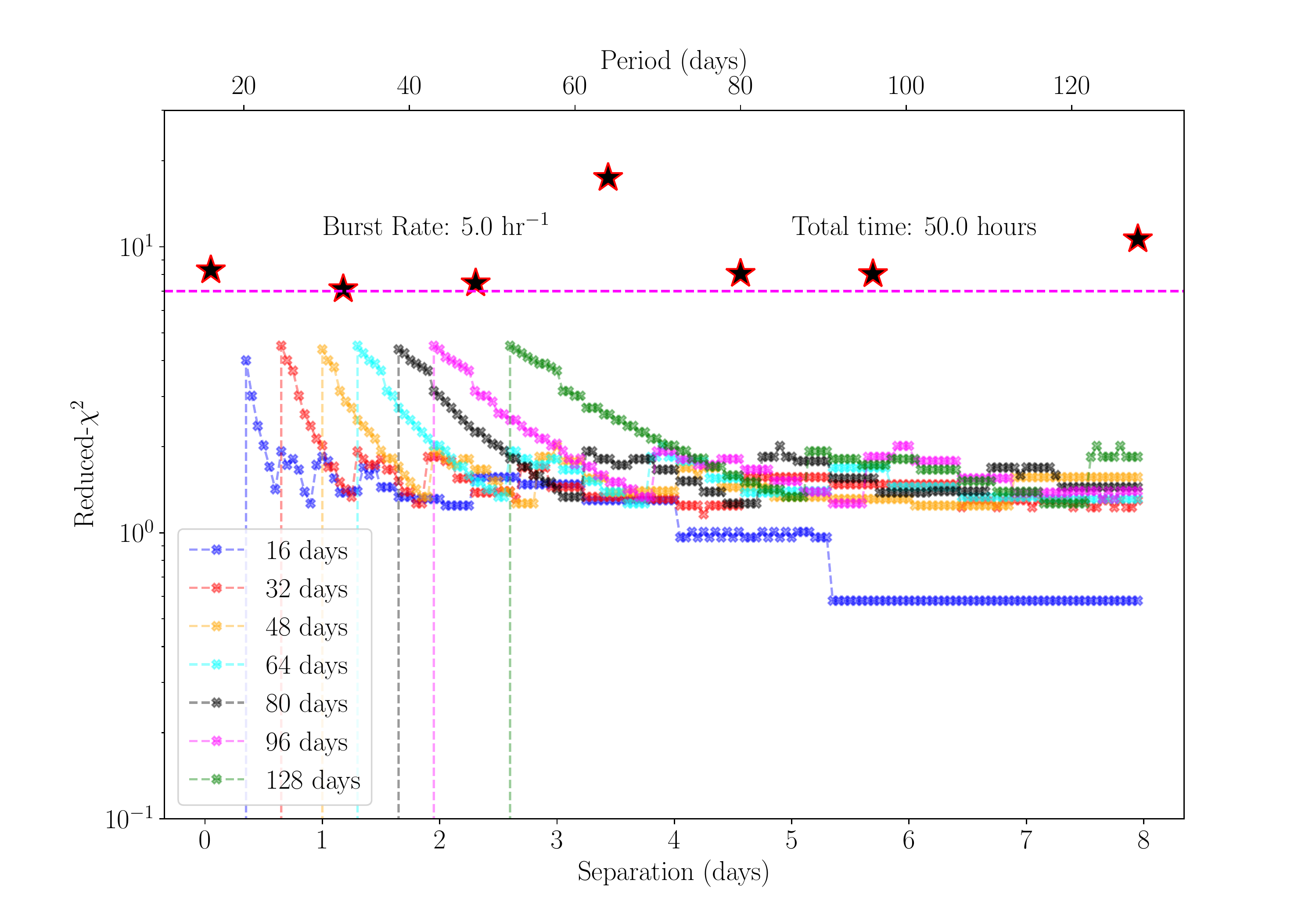}}
\caption{Reduced $\chi^{2}$ as a function of separation between observations for 50 hours of total allocated telescope time. Each panel corresponds to a different assumed burst rate as displayed in each panel. Different lines correspond to FRBs with different periods shown in the legend. We assume a duty cycle of 50$\%$ for the activity cycle. The black stars with red outlines correspond to the reduced-$\chi^{2}$ values obtained by CHIME for different periods (x-axis on the top of the panels) for a separation of 1 day and a source transit time of 15 minutes. The dashed magenta line corresponds to the reduced $\chi^{2}$ corresponding to a 5-$\sigma$ detection of the periodicity. The vertical dashed lines correspond to the minimum separation before a pulse is detected.}
\label{fig:simres1}
\end{figure*}

\begin{figure*}
\subfloat{\includegraphics[scale=0.33]{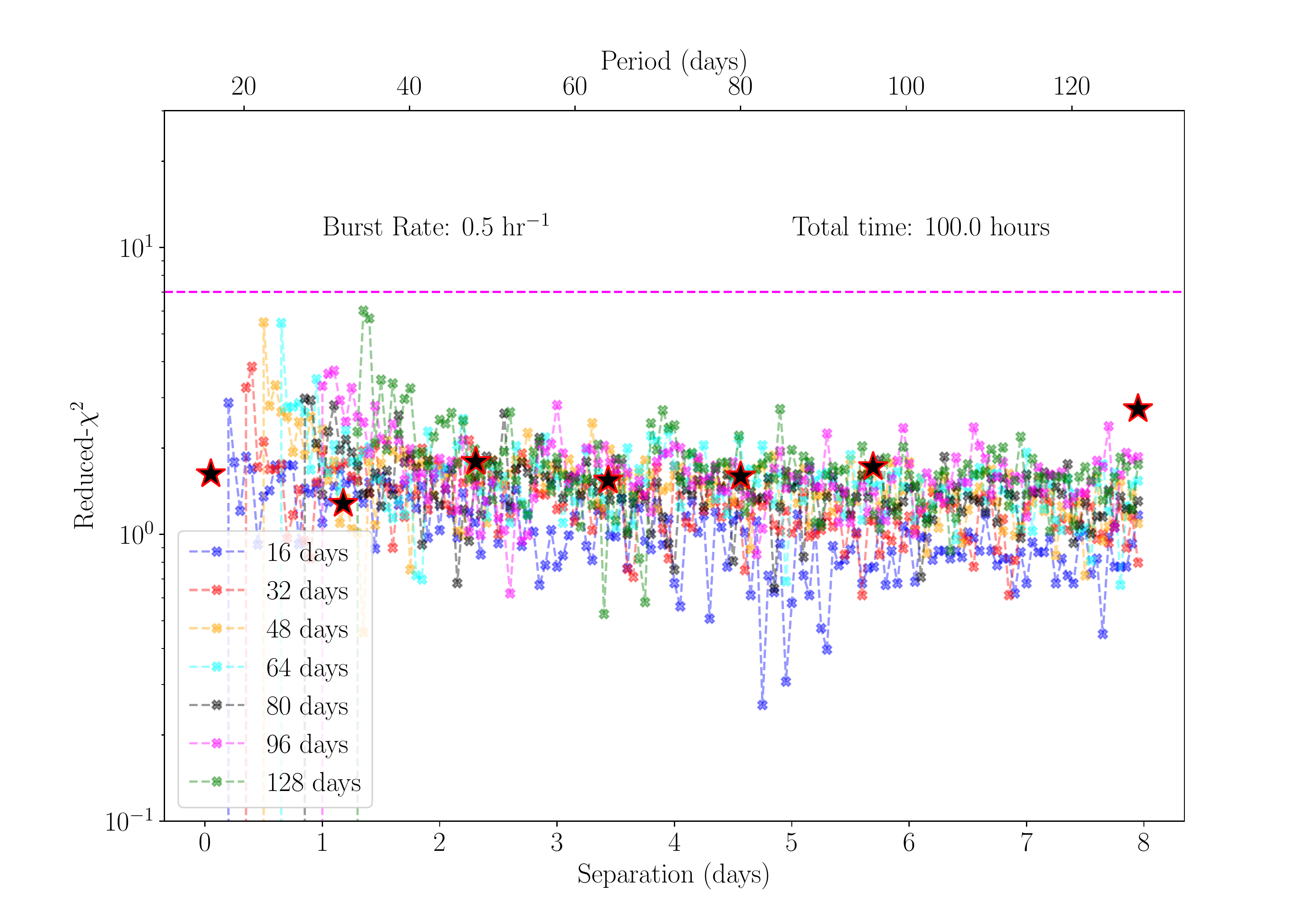}}
\subfloat{\includegraphics[scale=0.33]{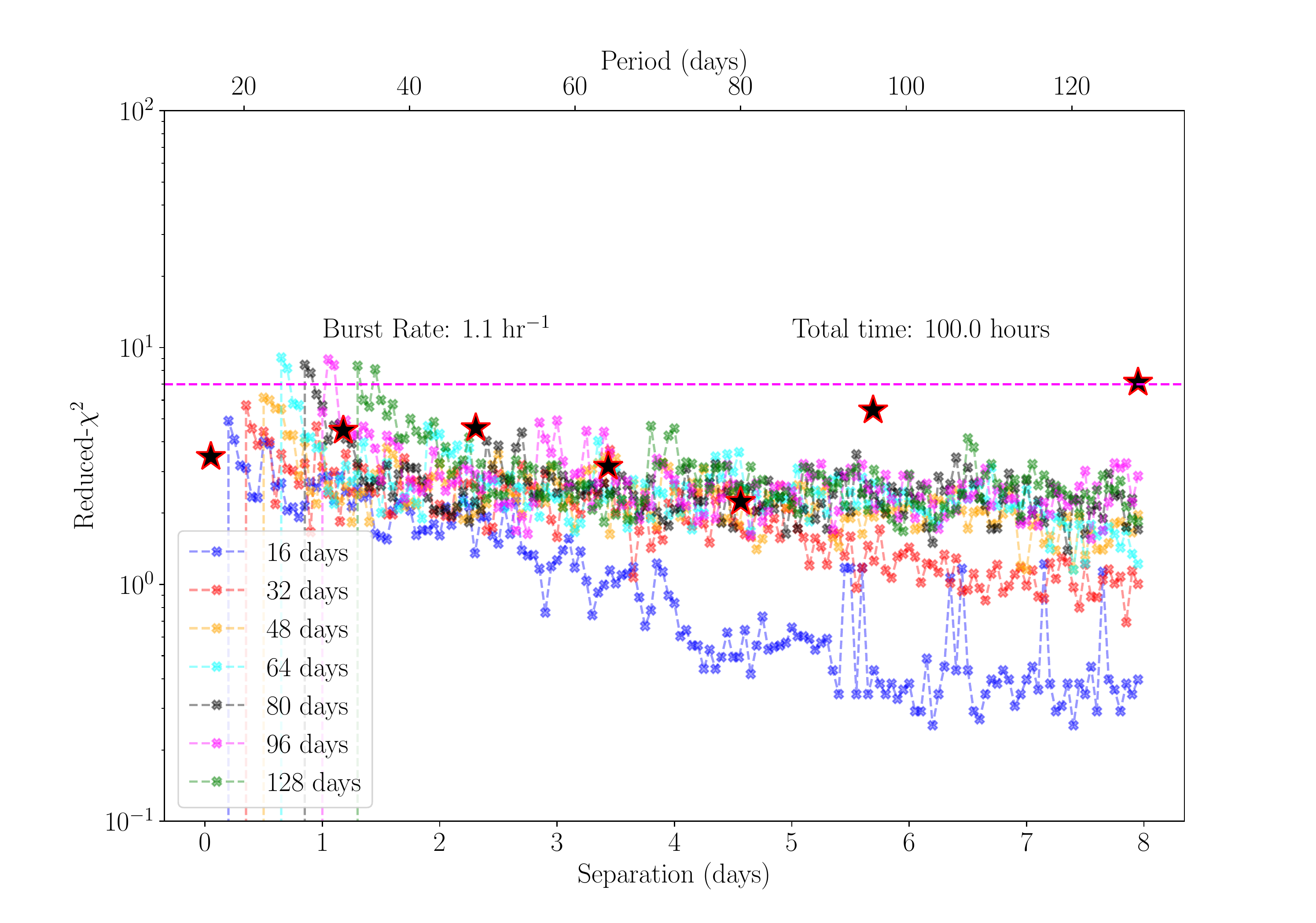}}\\ 
\subfloat{\includegraphics[scale=0.33]{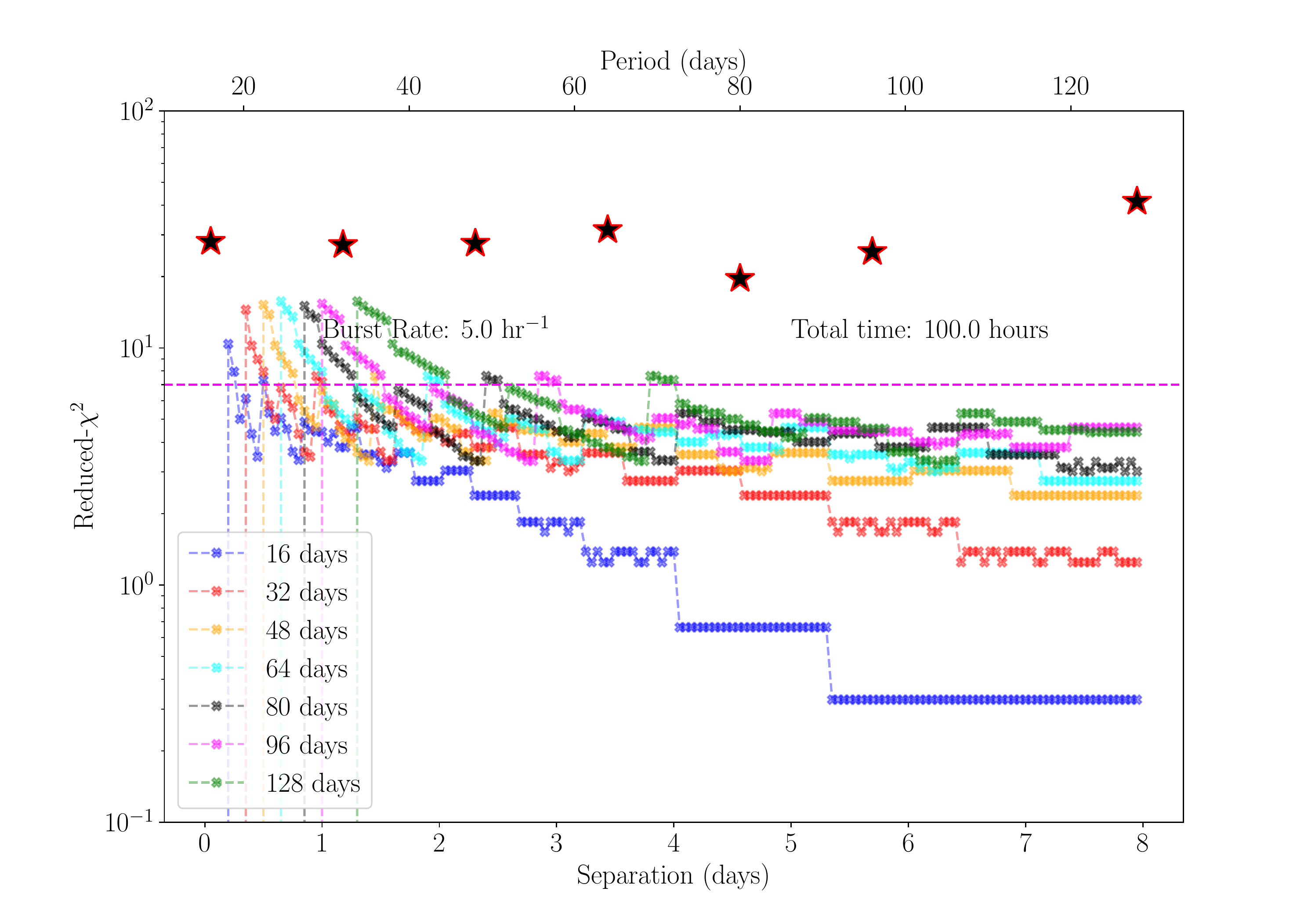}}
\caption{Reduced $\chi^{2}$ as a function of separation between observations for 100 hours of total allocated telescope time. Each panel corresponds to a different assumed burst rate as displayed in each panel. Different lines correspond to FRBs with different periods shown in the legend. We assume a duty cycle of 50$\%$ for the activity cycle. The black stars with red outlines correspond to the reduced-$\chi^{2}$ values obtained by CHIME for different periods (x-axis on the top of the panels) for a separation of 1 day and a source transit time of 15 minutes. The dashed magenta line corresponds to the reduced $\chi^{2}$ corresponding to a 5-$\sigma$ detection of the periodicity. The vertical dashed lines correspond to the minimum separation before a pulse is detected.}
\label{fig:simres2}
\end{figure*}

The analysis of FRB~121102 detections begs the question of whether all repeating sources of FRBs exhibit periodic activity. If we assume this to be the case, it has implications on follow-up strategies of future discoveries of repeating FRBs. We note that transit instruments such as CHIME will have an advantage over other steerable radio telescopes as transit instruments will automatically get a cadence of one day as the source transits in the beam of the telescope. Though transit telescopes can suffer from aliasing due to the fixed cadence of observations, they will be vital in discovering periodicities in repeating FRBs as any repeating FRBs with activity periods much larger than a day will not be affected by the window function.
In spite of this advantage, it is possible to get an optimized follow-up strategy for other single dish telescopes and interferometers. To that end, we ran a simulation to optimize follow-up strategies of periodic FRBs. To make our simulations agnostic to different observatories and different sensitivities, we assign unity weight to all observations where we detect a pulse and zero weight when there is a non-detection. In these simulations, we assume the FRB is emitting for 50$\%$ of the activity period. We assume that the bursts follow a Poissonian distribution in the active phase with a repetition rate of 1.1 bursts per hour at 1.4~GHz~\citep{houben2019}. We also ran the simulation for repetition rates of 0.5 and 5 bursts per hour to assess the effect of the burst rate on the detectability of a periodicity. During an active phase, for each observing session, we draw from a binomial probability distribution to check if a pulse was detected. The probability of detecting $N$ bursts for a given observing session of duration $T_{\rm obs}$,
\begin{equation}
P(X = N) = \frac{(R { T_{\rm obs}})^{N}~e^{-R{ T_{\rm obs}}}}{N!},
\end{equation}
where, $R$ is the repetition rate. Hence, the probability to detect any $N > 0$, $P(N > 0) = 1- e^{-R T_{\rm obs}}$. We use this computed probability to draw from the binomial distribution
to get the number of observing sessions within the activity period where there was a detection. This way, we take into account the sporadic nature of FRBs during an active period. Then, for a given activity period, we can obtain a sequence of detections and non-detections for our follow-up campaign over a range of separations between observations. 

For the follow-up campaign, we assume that each observing session is one hour long. Then, we assess the significance of the true periodicity that is obtained from the simulated detections. In order to achieve this, we generated a folded profile from the obtained sequence of detections and ran a goodness of fit test on it for a null hypothesis that the folded profile is uniform across the entire period. We use the reduced $\chi^{2}$ as the test statistic to evaluate the deviation of the resulting profile from the null hypothesis. We note that there is an underlying assumption here that all events within a phase bin of the folded profile follow Gaussian statistics which may not necessarily be true~\citep[see][for more details]{CHIME2020a}. We use a reduced $\chi^{2}$ of 7.0 as a threshold for the detection of a period at a 5-$\sigma$ level of significance after taking into account the number of trial periods searched in a putative FFA search. Since time on a telescope for such follow-up observations is limited, we ran this analysis for different amounts of allocated time on a any given radio telescope. Figures~\ref{fig:simres1} and~\ref{fig:simres2} shows the reduced $\chi^{2}$ as a function of separation of observations for 50 and 100 hours of observing time. The results of the simulation clearly show that for a burst rate of 0.5 hr$^{-1}$ and FRB~121102 like sources, one would need more than 100 hours of observing time to detect a significant period in the range of 10--150 days. On the other hand, CHIME will be able to detect repeaters with higher burst rates within 50 hours of on source time while other single dish telescopes will need at least 100 hours to detect high burst rate sources. Also, it shows that in order to obtain an accurate and significant detection of periodicity, one needs to have a fairly dense cadence of observations. While CHIME has the advantage of daily cadence, targeted follow-up campaigns will need a cadence ranging from 0.5--3 days in order to have the best chance to detect a period regardless of the burst rate.

\section{Conclusions}
We have carried out a long-term radio monitoring campaign of FRB~121102 with the Lovell Telescope. Using these pulses and other detections from the literature, we performed a periodicity search and detected a tentative period of 157 days in the periodogram with a duty cycle of 56$\%$. We extrapolated the computed period to the most recent activity and show that the detections lie within the activity phase predicted by the period. We do note that the uneven observing strategy prevents us from determining a robust significance of the detection of the said period. To avoid these issues in the future, we performed simulations of periodic FRBs to show that non-transit telescopes need at least 100 hours of follow-up time to determine periodicities in these sources. This shows that single dish telescopes and interferometers will be able to follow-up repeating FRBs in reasonable amount of telescope time to detect periodicities. Our study also shows the importance of reporting non-detections for any repeating FRB follow-up campaigns as they are crucial for computing the robustness of any detected periodicity. If the periodicity in FRB 121102 is genuine, it suggests that there is a large range in the periodicities of repeating FRBs and more periodic FRBs need to be discovered to infer the nature of their progenitors.

\section*{Acknowledgements}
We would like to thank our anonymous reviewer whose remarks vastly improved the manuscript. KMR, BWS, VM and MC acknowledge funding from the European Research Council (ERC) under the European Union's Horizon 2020 research and innovation programme (grant agreement No 694745). DA and DRL acknowledge support from the National Science Foundation awards AAG-1616042, OIA-1458952 and PHY-1430284. RPB acknowledges support from the European Research Council under the European Union's Horizon 2020 research and innovation programme (grant agreement no. 715051; Spiders). The authors would like to thank Andrew Lyne for help with the Lovell Telescope observations during this campaign.




\bibliographystyle{mnras}
\bibliography{refs} 



\appendix


\bsp	
\label{lastpage}
\end{document}